\documentclass[pra,preprint,amssymb,showpacs]{revtex4}
\usepackage{dcolumn}
\usepackage{bm}
\usepackage{graphicx}
\def\be{\begin{equation}}
\def\ee{\end{equation}}
\def\bea{\begin{eqnarray}}
\def\eea{\end{eqnarray}}
\def\source#1#2#3#4{{\it #1}~{\bf #2}, #3 (#4)}
\def\Eq#1{Eq. \ref{#1}}
\def\Eqs#1#2{Eqs. \ref{#1} and \ref{#2}}
\def\Ref#1{Ref. \cite{#1}}
\def\Refs#1#2{Refs. \cite{#1} and \cite{#2}}

\def\ird{\frac{1}{\sqrt{d}}}

\def\ket#1{| #1 \rangle}

\def\bra#1{\langle #1 |}
\def\braket#1#2{\langle #1| #2 \rangle}

\def\ave#1{\langle \Psi | #1 | \Psi \rangle}
\def\ketPsi{|\Psi\rangle}

\def\phio{\phi_o}

\def\PsiPhi{|\Psi(\Phi)\rangle}

\def\om{\omega}

\def\SO2N{SO(2)^{\otimes N}}
\def\hbe{\hbox{e}}

\def\hz{\hat{z}}
\def\hzk{\hat{z}_k}

\def\phik{\phi_k}
\def\setphik{\{ \phi_k \}}
\def\BRphik{\BR(\{\phik\})}
\def\BRinvphik{\BR^{-1}(\{\phik\})}
\def\tY{\tilde{Y}}
\def\tD{\tilde{\Delta}}
\def\Del#1{\Delta^{(#1)}}
\def\YY#1{Y^{(#1)}}
\def\BX{{\bf X}}

\def\BR{{\bf R}}
\def\BXphik{{\bf X}(\setphik)}
\def\BXPhi{{\bf X}(\Phi)}

\def\CT{{\cal T}}
\def\CTN{{\cal T}^{\otimes N}}
\def\CO{{\cal O}}

\def\cx{{\cal X}}
\def\cy{{\cal Y}}
\def\cxk{{\cal X}_k}
\def\cyk{{\cal Y}_k}
\def\SO2N{SO(2)$^{\otimes N}$}
\begin{document}
\title{Rotational Covariance and Greenberger-Horne-Zeilinger theorems for three or more particles of any dimension}
\author{Jay Lawrence} 
\affiliation{Department of Physics and Astronomy, Dartmouth
          College, Hanover, NH 03755, USA}
\affiliation{The James Franck Institute, University of Chicago, 
          Chicago, IL 60637}
\date{revised \today}
\bigskip
\begin{abstract}
Greenberger-Horne-Zeilinger (GHZ) states are characterized by their transformation properties 
under a continuous symmetry group, and $N$-body operators that transform covariantly 
exhibit a wealth of GHZ contradictions.  We show that local or noncontextual hidden variables 
cannot duplicate the predicted measurement outcomes for covariant transformations, and we 
extract specific GHZ contradictions from discrete subgroups, with no restrictions on particle 
number $N$ or dimension $d$ except for the general requirement that $N \geq 3$ for GHZ 
states.  However, the specific contradictions fall into three regimes distinguished by 
increasing demands on the number of measurement operators required for the proofs.   
The first regime consists of proofs found recently by Ryu et. al. \cite{RLZL}, the first 
operator-based theorems for all odd dimensions, $d$, covering many (but not all) particle 
numbers $N$ for each $d$.   We introduce new methods of proof that define second and
third regimes and produce new theorems that fill all remaining gaps down to $N=3$, for 
every $d$.   The common origin of all such GHZ contradictions is that the GHZ states and 
measurement operators transform according to different representations of the symmetry 
group, which has an intuitive physical interpretation.  
\end{abstract}
\maketitle
\section{Introduction}

The groundbreaking discovery by Greenberger, Horne and Zeilinger \cite{GHZ,GHZS} of 
nonprobabilistic contradictions \cite{HR83} between quantum predictions and local hidden 
variables theories (now called GHZ contradictions), and Mermin's demonstration \cite{Mermin1} that 
these embody Kochen-Specker contextuality \cite{Bell1966,KS} as well as Bell-EPR nonlocality 
\cite{Bell1964}, sparked vigorous developments along a number of lines.  These included GHZ 
theorems for many qubits \cite{Mermin2,Ardehali}, improved Bell \cite{Hardy} and Kochen-Specker 
theorems \cite{KP,Cabello96,Lisonek}, Bell theorems for two particles of arbitrary dimension $d$ 
\cite{Kas.et.al.,Collins,Fu}, and the introduction of noncontextuality inequalities, state dependent 
\cite{Klyachko} and state-independent \cite{Cabello08,Badziag}. 
Connections of GHZ theorems to practical pursuits of quantum cryptography \cite{QSS,SG01} and 
quantum error correction \cite{DiVP} have been made.  An error correction protocol employing 
concatenated GHZ states in particular was recently proposed \cite{Frowis}. 


Of particular interest here is the extension of GHZ contradictions to systems of both higher dimension 
$d$ and a broader range of particle numbers $N$.  Zukowski and Kaszlikowski \cite{ZK99} described 
an experimental protocol involving spatially separated arrays of beam splitters, phase shifters, and 
detectors that would show GHZ contradictions for $N$ particles, each of dimension $d=N-1$.  The 
same authors \cite{KZ02} found a similar protocol for $N$ particles, each of dimension $d=N$, but 
with only probabilistic quantum predictions.  Cerf, Massar, and Pironio \cite{CMP} found GHZ 
theorems in the form of Kochen-Specker operator identities, based on a compatible set of Pauli 
operators (stabilizers) analogous to Mermin's, for all even dimensions $d$ and all odd $N \geq d+1$.   
They also established criteria for a contradiction to be genuinely  (or irreducibly ) multiparty 
($N$) or multidimensional ($d$).  Lee, Lee, and Kim  \cite{LLK} extended these results to include 
{\it all} odd $N \geq 3$, for every even $d$.   They accomplished this by using concurrent operators - 
operators which have a common eigenstate even if they do not commute.  They realized that a 
common eigenstate allows the establishment of EPR elements of reality \cite{EPR} and thus GHZ 
contradictions, although noncommutativity makes these contradictions state-dependent and rules 
out KS operator identities.   Recently, Tang et. al. further extended these results by deriving 
contradictions for all even $N \geq 4$, for all even $d$.   The $d \geq 4$ proofs were based on 
stabilizers of GHZ graph states, allowing state-independent contextuality inequalities \cite{TYO13}, 
while the more challenging $d=2$ proofs (with $N$ even) used concurrent operators with more than 
two measurement settings for each particle \cite{TYO13b}.   Waegell and Aravind \cite{WA13} have 
explored systematically both observable-based and projector-based proofs of the KS theorem, 
based on the $N$-qubit Pauli group for all numbers of qubits $N \geq 2$.   Such proofs number 
in the thousands, and a subset can be converted into GHZ paradoxes \cite{WA12} for all even 
$N \geq 4$.  Very recently, Ryu et. al. \cite{RLZL,RLZL2} found GHZ theorems for all $d$, with 
infinite sequences of $N$ for each $d$, answering a long-standing question whether odd-$d$ 
contradictions could be found beyond the line $N=d+1$ of \Ref{ZK99}.   They did this by extending 
the concurrent operator approach introduced in \Ref{LLK}, so that these contradictions are also 
state-dependent.  This result sharpened the question whether state-independent
contadictions could be found for odd $d$ using compatible sets of Pauli operators.   While deriving 
state-{\it de}pendent contextuality  proofs for such cases,  Howard et. al. \cite{Howard} answered the 
above question in the negative, arguing that previous results of Gross and of Veitch et. al. \cite{other} 
rule out state-independent proofs with stabilizer measurements of any odd $d$.

Also of interest for this work is the general connection of symmetry to GHZ contradictions.  It is 
noteworthy that the seminal GHZ paper \cite{GHZ} used the rotational symmetry of a four-particle 
state to derive contradictions; it was noted in passing that three particles would suffice.   The only 
other derivations that make explicit use of GHZ state symmetry, as far as I am aware, are the 
concurrent operator approaches \cite{LLK,TYO13b,RLZL,RLZL2}.   There, tensor product 
operators are found that preserve the GHZ state while transforming a particular $N$-body 
operator into others that form a concurrent set. 

In this work we suggest a broader role for symmetry arguments in deriving and interpreting GHZ 
contradictions. With GHZ states providing an appropriate first case, we describe an $N$-body uniaxial 
rotation group that characterizes GHZ entanglement and generates continuous sets of covariant 
$N$-body operators, of which concurrent operators form a subset.  The continuous group properties 
are used in a simple proof that local or noncontextual hidden variables (HVs for short) are incapable 
of replicating predicted measurement outcomes under covariant transformations.  This proof places no 
restrictions on dimension or particle number except for the general requirement $N \geq 3$ for GHZ
states \cite{Ngeq3}.  This raises the question of whether specific, experimentally accessible GHZ 
contradictions can be found using discrete subgroups for the same cases:  Are the existing gaps 
fundamental or technical in origin?  To answer this question, we present a succession of three methods
of proof that together succeed in filling all the gaps.  This succession reveals three distinct regimes 
of specific contradictions, each more demanding than its predecessor in terms of their requirements
for both the number of concurrent $N$-body operators, and the number of one-qudit measurement 
bases.  We attempt to minimize both numbers in order to provide the simplest protocols for 
experimental tests at each level.
%

In the next section we describe the characteristic rotational symmetry of GHZ states, the covariance of 
operators, and the identification of concurrent operators.   We conclude the section with a formal proof 
that hidden variables cannot replicate the covariance of operators.  In Sec.  3  we derive the three 
regimes of specific GHZ contradictions, completing the catalog of all $N \geq 3$ for every $d$.   We 
conclude this section with a physical interpretation of the contradictions based on the covariance.   In 
the concluding Sec. 4, we summarize the results and discuss remaining open questions.
 
\section{Rotational Symmetry}

A generalized GHZ state of $N$ qudits can be written as 
\be
  \ket{\Psi}  = \ird \sum_{n=0}^{d-1} \ket{nn...n},
\label{GHZ1}
\ee
where the indices refer to the eigenvalues of the one-qudit Pauli $Z$ operators,
\be
   Z = \sum_{n=0}^{d-1} \ket{n} \om^n \bra{n},  \hskip1truecm  \hbox{where}
   \hskip1truecm  \om \equiv \exp(2\pi i/d).
\label{Zop}
\ee
Imagining the qudits as $d$-component spinors, $Z$ is proportional to an exponential of the 
spin operator, $S_z$, whose eigenvalues are $m = S, S-1, ..., -S$, with $2S+1= d$.  The spectra 
of $Z$ and $S_z$ are related by $n = S-m$, so the operators are related by $Z = \om^{(SI-S_z)}$,  
and rotations about $\hz$ axes may be written as powers of $Z$,
 \be
     R(\phi) = \exp (-i S_z \phi) =  \hbox{e}^{-i S \phi} Z^{\phi d/2 \pi}.
\label{Rotop}
\ee
Rotations of all $N$ qudits through independent angles $\phik$  about their 
respective $\hzk$ axes ($k=1,...,N$), are products of the individual rotations,
$\BR(\setphik) = \prod_k R_k(\phi_k)$.  These rotations form the group of $N$ independent 
uniaxial rotations, $\CTN$,  where $\CT$ is the group of rotations in a plane, or the circle group.
Applied to the state $\ket{\Psi}$ above, we find 
\be
   \BR(\setphik) \ket{\Psi} = \ird \hbe^{-iS \Phi} 
   \sum_{n=0}^{d-1} \hbe^{i n \Phi} \ket{nn...n} \equiv  \PsiPhi,  \hskip1.3truecm
     \Phi = \sum_{k=1}^N \phik,
\label{RotPsi}
\ee
so that the transformed state depends only on the compound (net) rotation angle $\Phi$, and
reveals nothing about the individual $\phik$.   Because of this, we can represent $\PsiPhi$ 
uniquely on the unit circle.  Moving around from 0 to $2\pi$, one encounters $d$ orthogonal 
states at integral multiples of $2\pi /d$, as illustrated for the case of $d=3$ in Fig. 1(a).  To show 
this orthogonality, note that the inner product of any two states on the unit circle is 
\be
   \braket{\Psi(\Phi')}{\Psi(\Phi)} = \bra{\Psi} R(\Phi - \Phi') \ket{\Psi} 
   =  \frac{\sin d(\Phi - \Phi')/2}{d \sin (\Phi - \Phi')/2}.
\label{inner}
\ee
This vanishes if and only if the two angles differ by a nonzero multiple of $2\pi/d$ and confirms 
that the states $\ket{\Psi(2\pi\nu/d)}$, for $\nu=0,1,...,d-1$, form an orthonormal set.  Thus, $\ket{\Psi}$ 
transforms as a $d$-dimensional representation of $\CT$.  Note in passing that \Eq{RotPsi} signals a 
sign change under any compound $2\pi$ rotation for systems composed of half-integral spin 
(even-$d$) particles.  

\begin{figure}[h!]
\includegraphics[scale=0.60]{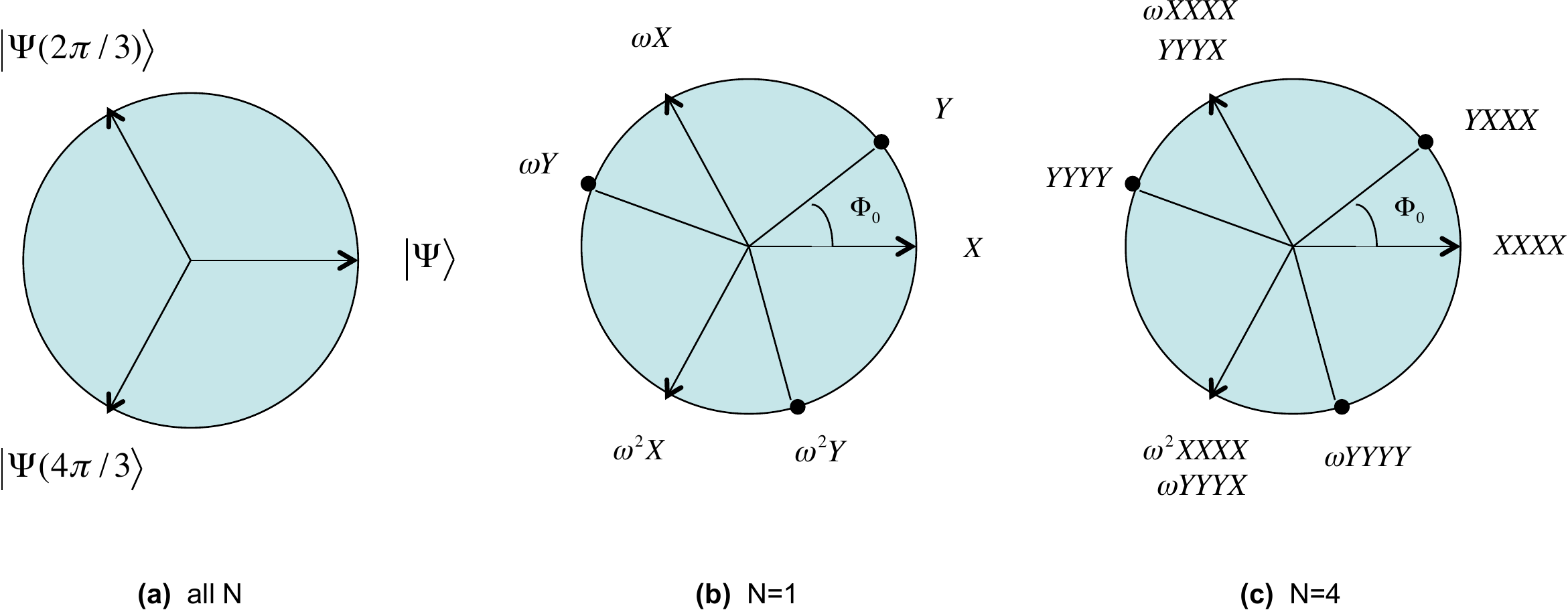}
\caption{\label{fig1} Circle plots for $N$ qutrits ($d=3$): (a) orthonormal set of GHZ states for any 
$N$, (b) periodicity property of operators for one qutrit,  and (c) periodicity property of tensor product 
operators for 4 qutrits.  One locates tensor product operators on the plot by adding up the angular 
variables of all the factors.}
\end{figure}

Now consider the rotational properties of operators.  The crucial operator of which $\ketPsi$ 
is an eigenstate (with eigenvalue unity) is the tensor product,
\be
   \BX = \prod_k X_k,
\label{bigX}
\ee
where the one-qudit operators $X_k$ are the usual raising operators of $Z_k$,
\be 
   X_k = \sum_{n=0}^{d-1} \ket{n+1} \bra{n},
\label{xop}
\ee
with the convention $\ket{d} \equiv \ket{0}$ understood.   We shall refer to $\BX$ and the
$X_k$ as observables because (like $Z_k$) they are unitary, and therefore exponentials of 
Hermitian operators whose eigenvalues are $0, 1, ..., d-1$.
We define covariant rotations of $\BX$ as those which preserve eigenvalue relations in 
rotated states,
\be
   \BRphik \BX \BRinvphik   \equiv  \BXphik    = 
    \prod_k X_k(\phik),
\label{RotBX1}
\ee
so that $\PsiPhi$ is an eigenstate of any $\BXphik$ for which $\sum \phik = \Phi$.   More generally,
considering relative rotations of operator and state, the expectation value 
$\ave{ \BRphik \BX \BRinvphik}$ reflects the fact that rotating {\it only} the operator has the same 
effect on measurement outcomes as rotating only the state in the opposite sense.

The rotated one-qudit factors in \Eq{RotBX1} are given by
\be
    X_k(\phik) = \hbe^{i\phik} \sum_{n=0}^{d-2} \ket{n+1}\bra{n}
   + \hbe^{i(1-d)\phik} \ket{0} \bra{d-1}.
\label{rotXk1}
\ee
Each factor $X_k(\phik)$ has two key properties:  (i) It is e$^{i\phik}$ times a periodic function of 
$\phik$ with period $2\pi/d$:
\be
   X_k(\phik + 2\pi/d) = \hbe^{2\pi i/d} X_k(\phik) \equiv \om X_k(\phik),
\label{rotXk2}
\ee
and (ii) it lives in a 2-dimensional operator space spanned by $X_k$ and another operator, 
$Y_k$, which may be defined at an arbitrary point within the first period,
\be
  Y_k \equiv X_k(\phio),  \hskip1.3truecm 0 < \phio < 2\pi/d.
\label{Yop}
\ee
The periodicity is illustrated in Fig. 1(b) for the case of $d=3$, where the underlying period 
is $2\pi/3$, and $Y$ is chosen to be $X(2\pi/9)$ for the purpose of Sec. III.   For $d > 2$, $X$ and 
$Y$ form a nonorthogonal operator basis (whatever the choice of $\phio$).   For $d=2$ (not shown), 
the choice $Y=X(\pi/2)$ is the usual Pauli matrix, producing an orthogonal basis, [Tr($XY)=0$].

There are infinitely many rotated $N$-qudit operators $\BXphik$ that correspond to a
particular collective angle $\Phi$.   It is useful to denote the set of all such operators as $[\BXPhi]$: 
\be
   [\BXPhi]  \equiv  \bigg\{ \BXphik: \sum_{k=0}^{d-1} \phik = \Phi \bigg\},
\label{setBX}
\ee
and to associate this set with the point $\Phi$ on the unit circle.   Examples of discrete subsets 
consisting of factors of just $X$s and $Y$s are given in Fig. 1(c).   Now, because an element of 
the set $[\BX(\Phi)]$ is generated by rotating a single qudit through $\Phi$, this set must have the 
same periodicity property (\Eq{rotXk2}) as an individual $X_k$, that is, $[\BX(\Phi + 2\pi/d)] = \om 
[\BXPhi]$.  Therefore, at the special points $\Phi = 2\pi\nu/d$ ($\nu = 0,1,...,d-1$) on the unit circle, 
we have
\be
   [\BX(2\pi\nu/d)] = \om^{\nu} [\BX].
\label{rotBX3}
\ee
Recalling that $\ket{\Psi}$ (\Eq{GHZ1}) is an eigenstate of $\BX$ with eigenvalue 
unity, \Eq{rotBX3} shows that it is also an eigenstate of every operator in the set 
$[\BX(2\pi\nu/d)]$ with eigenvalue $\om^{\nu}$,
\be
        [\BX(2\pi\nu/d)] \ket{\Psi} = \om^{\nu} \ket{\Psi}.
\label{eigenvalues3}
\ee
Covariance implies that the GHZ state at {\it any} special point, $\ket{\Psi(2\pi\mu/d)}$, is an 
eigenstate of operators  at $\nu$ with eigenvalue $\om^{\nu - \mu}$.     In terms of concurrency, 
one could say that the set of operators $\cup_{\nu = 0}^{d-1} [\BX(2\pi\nu/d)]$ is concurrent with 
respect to the set of GHZ states $\{\ket{\Psi(2\pi\mu/d)}: \mu = 0, 1, ..., d-1\}$.  For the special case 
of $d=2$ with the choice of Pauli operators, the corresponding set is also compatible.

Equation \ref{eigenvalues3} forms the basis of specific GHZ contradictions, and readers who so
desire can skip to Sec. III where that discussion begins, or continue with the formal proof 
immediatetly following, which may be useful for later developments.

\vskip0.3truecm
\centerline{{\bf General Failure of Hidden Variables}}
\vskip0.2truecm

The assumption embodying local realism, or more generally noncontextuality, is that each one-qudit 
factor takes a definite value, $v(X_k(\phi))$.   Since this value must be an eigenvalue of $X_k(\phi)$, 
it is natural to parameterize it as
\be
   v(X_k(\phi)) = \om^{\cxk(\phi)},
\label{LHV1}
\ee
where $\cxk(\phi) = 0,1,...,d-1$.   To conform to $N$-body measurements as described by quantum 
theory, these variables must be individually random, but correlated so as to reproduce definite 
$N$-body products when the GHZ state under consideration is an eigenstate.  Here we shall 
demonstrate, in general terms, that such variables are incapable of replicating the covariance 
described above.   We shall proceed by  demanding that they respect the invariance of 
measurement outcomes (eigenvalues) under any $\Phi$-preserving rotations (we will call this GHZ 
{\it in}variance), and then show that they fail to transform covariantly under any $\Phi$-changing 
rotation.   

To impose GHZ invariance, we shall consider measurements on the state $\ket{\Psi}$ corresponding 
to several operators at $\nu = 0$, all of which produce the value 1 with certainty.   We choose $\BX$ 
and any operator obtained from it by rotating two qudits through opposite angles, namely 
$\CO_{ij} = [X_i(\phi)X_i^{-1}X_j(-\phi)X_j^{-1}]\BX$.     Equating the values, $v(\CO_{ij}) = v(\BX)$, 
relates the ratios of individual factors, $eg$, $v(X_i(\phi))/v(X_i)$, which depend only on the 
{\it variations} of the exponents with angle, $\Delta \cx_i(\phi) \equiv  \cx_i(\phi) - \cx_i(0)$.  The 
resulting equation is
\be 
   \Delta \cx_i(\phi) + \Delta \cx_j(-\phi) = 0.
\label{LHV2}
\ee 
Since this equation applies to all $i$ and $j$, the variations are uniform over qudits,
\be
   \Delta \cx_1(\phi) = ... = \Delta \cx_N(\phi) \equiv \Delta \cx(\phi), 
\label{LHV3}
\ee
that is, there is a single variation, and this is an odd function of $\phi$,
\be
   \Delta \cx(\phi) = -\Delta \cx(-\phi).
\label{LHV4}
\ee
The variations are thus constrained even though the functions $\cx_i(\phi)$ themselves are random. 

Assume now that we have $N \geq 3$ qudits, divide them into two unequal groups consisting of 
$N_1$ and $N_2$ qudits each, and let $\lambda = N_2/N_1 > 1$.   Rotate the first group through 
$\phi$, and the second group through ($- \phi/\lambda$), making $\Phi = 0$.   Using 
\Eqs{LHV3}{LHV4}, we arrive at
\be
   \Delta \cx(\phi) = \lambda \Delta \cx(\phi/\lambda).
\label{LHV5}
\ee
This equation requires that $\Delta \cx(\phi)$ be a linear function of $\phi$.   But since it can only take 
discrete values, it must be a constant, and by definition that constant must be zero.   So 
$\Delta \cx(\phi) = 0$ for every qudit.  This means that while every $\cx_k(\phi)$ is a random variable, 
it must be isotropic.  Together, these $\cx_k(\phi)$ predict that  the $N$-body values $v(\BX(\setphik))$ 
cannot vary with $\Phi$, and, in the state $\ket{\Psi}$, all are unity.  If instead we chose the state 
$\ket{\Psi(2\pi \nu/d)}$ for this proof, the predicted value would be $\om^{\nu}$.   In short, hidden 
variables that respect GHZ invariance must fail the covariance.   


It is interesting that this general failure is proven independently of $d$ and $N$, except for the 
condition that $N \geq 3$.  Unfortunately, this result does not lead immediately to GHZ contradictions
that are congenial to experimental tests.   While one can test the invariance of eigenvalues of $\CO_{ij}$
at particular choices of $\phi$, thus establishing that $\cx_k(\phi)$ is an element of reality for those 
choices, this does not prove  it for every $\phi$.   Testable contradictions may be found in this approach 
by identifying an appropriate angular interval ($\phio$) that defines a finite set of individual factors 
($X$ and $Y$ in the simplest case),  and constructing a finite (minimal) set of $N$-body tensor 
products for the proof.  An experimental test requires measurements of all these, and each in turn
requires individual qudit measurements in bases dictated by the factors.   Since the number of
such factors must be kept finite, less can be inferred about individual 
hidden variable properties than in the proof above.   In fact, in existing proofs such inferences need 
not be made explicitly, although they are present implicitly.   In the succession of proofs to follow, we 
shall see that as more operators are required, more inferences become possible, and that at some 
level the proofs may be simplified by making these inferences explicit.

\section{GHZ Contradictions}

One can identify three regimes of GHZ contradictions according to increasing numbers of measurements
required for experimental tests.  In this section we present three methods of proof that define these 
regimes.  The first proof is similar to that of Ryu et. al. \cite{RLZL}; the contradictions are equivalent and 
the values of $N$ and $d$ are the same.   While it is possible to infer HV properties from the concurrent
operators used, the proof is simpler without this.
 
\vskip0.3truecm
\centerline{{\bf Method 1}}
\vskip0.2truecm

We choose several tensor products of $X$s and $Y$s at $\nu=1$ (as in Fig. 1c) and obtain the 
contradiction by comparing with $\BX$ (at $\nu=0$).  To appropriately define the $Y$ factors for 
the tensor products,  let $f$ be any factor of $d$ ($f \neq 1$), and choose  $\phio = 2 \pi / fd$, 
that is,
\be
   Y_k \equiv X_k(2\pi/fd),
  \label{Ychoice1}
\ee
which divides the basic period into $f$ parts.   (Fig. 1c corresponds to $f = d = 3$, and $N = 4$.)
Then, for any $N > f$, we find many operators at $\nu = 1$ that have $f$ factors of $Y$ and $N-f$ 
factors of $X$.  We need $N$ such operators, and we select those which have the $X$ and $Y$ 
factors grouped together,  as in $XXXYY$, $YXXXY$, etc., regarding $Y_1$ and $Y_N$ as neighbors.   

We now introduce hidden variables for the $X$ and $Y$ factors,  
\be 
   v(X_k) = \om^{\cxk}  \hskip1truecm  \hbox{and} 
   \hskip1truecm v(Y_k) = \om^{\cyk},
\label{LR1}
\ee
with $\cxk$ and $\cyk = 1,2,...,d-1$, and compare the requirements at $\nu= 0$ and 1:   Assuming 
that the system is in state $\ket{\Phi}$, in which $\BX$ has eigenvalue unity, we have 
\be
   \sum_k \cxk = 0.
\label{LR2}
\ee
The other $N$ operators all have eigenvalue $\om$ in $\ketPsi$.   Rewriting these eigenvalue 
equations in terms of $\cxk$ and $\cyk$ and then adding them all together, 
we find
\be
   (N-f) \sum_k \cxk + f \sum_k \cyk =  N.
\label{LR3}
\ee
Combining \ref{LR2} and \ref{LR3} leaves us with
\be
   f \sum_k \cyk = N,
\label{LR4}
\ee
with equality understood modulo $d$.   This equation has solutions only if $N$ is a multiple of $f$.   
Therefore, we have GHZ contradictions for all $N > f$ that are {\it not} multiples of $f$.  

If $d$ is a prime, then the above applies simply to $f=d$, and contradictions are found as shown in
the prime-$d$ columns of Fig. 2, with $N$-values denoted by red squares.

If $d$ has multiple factors, then the above proof applies to all of them, using different $\phio$ 
values for each.   Any one of these options suffices to establish a contradiction.  Therefore,
contradictions begin with $N = p_1+1$, where $p_1$ is the smallest (prime) 
factor of $d$, and above this they occur for every $N$ that is not a multiple of {\it every} factor of 
$d$ below $N$.  These contradictions are also represented by the red squares in Fig. 2.  The 
remarkable case of $d=12$ is understood in terms of the factors 2, 3, and 4. 

\begin{figure}
\includegraphics[scale=.60]{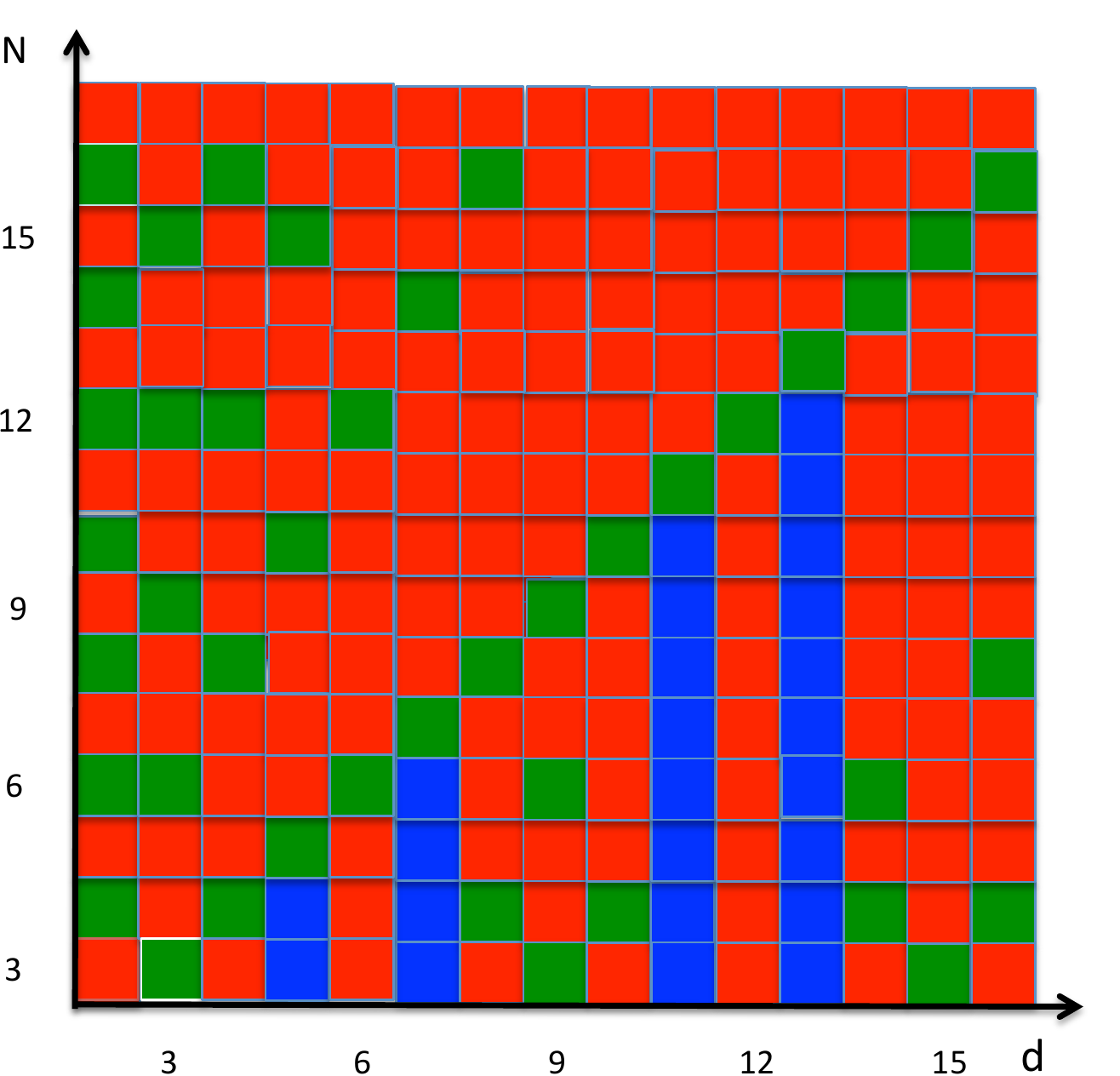}
\caption{\label{fig1} Red indicates known contradictions (\Refs{LLK}{RLZL}) recovered by method 1.
Green and blue indicate contradictions found by methods 2 and 3, respectively.} 
\end{figure}

One finds many other specific GHZ contradictions with this method, but I have found none with further 
$N$ values for any $d$.    To illustrate, consider examples with $d=3$:  When $N=4$, Fig. 1(c) shows a 
trivial alternative to the standard construction using operators at $40^o$ and $160^o$.   When $N=5$, 
a similar trivial alternative occurs at $80^o$ and $200^o$; less trivially, at $40^o$ and $160^o$, one 
finds operators of the type $YXXXX$ and $YYYYX$ respectively.   These provide five HV equations 
at each point, clearly different from \Eqs{LR2}{LR3}, but their combination reduces to the same final 
condition \ref{LR4}.   In the more interesting case of $N=6$, although we have the choice of five 
distinct sets of concurrent operators, the corresponding sets of equations all reduce to the standard 
one, which shows {\it no} GHZ contradictions.  Method 2 is complementary in providing proofs for just 
such cases.

\vskip0.3truecm
\centerline{{\bf Method 2:  Multiples of $d$ and its factors }}
\vskip0.2truecm

Here we derive GHZ contradictions, again for any $d$, where $N$ is a multiple of any factor of $d$.   
Regime 2 (the green squares) is defined as the set of all such cases not already assigned to regime 1 
(and colored red).    Using different methods, Refs. \cite{TYO13, TYO13b, RLZL2} have found 
contradictions for subsets of regime 2 \cite{subsets}.   In order to fill remaining gaps and provide a 
single derivation covering all of regime 2,  we introduce a new type of derivation which, in the spirit of 
the formal proof of Sec. 2, makes explicit reference to hidden variable failure.   

In analogy with the formal proof,  we define a conjugate of $Y$, namely $\tY \equiv X(-\phio)$, 
and use it to form a set of concurrent operators at $\nu = 0$, starting with
\be
 O_1 =   YX\hbox{...}X\tY,~ O_2 = XYX\hbox{...}X\tY, ~...,~ O_{N-1} = X\hbox{...}XY\tY,
\label{ConOps2a}
\ee
assuming that $N \geq 3$.  Equalities among the values $v(\CO_k)$ and $v(\BX)$ relate the ratios of 
individual qudit factors, which in turn depend only on variations in the exponents, defined here by
\bea
  &  \hbox{ln}_{\om}[v(Y_k)/v(X_k)] = \cyk - \cxk \equiv \Delta_k,  \\
  &  \hbox{ln}_{\om}[v(\tY_k)/v(X_k)] = \tilde{\cy}_k - \cxk \equiv \tilde{\Delta}_k.
\label{LHV1b}
\eea
From these equalities, we may deduce that
\be
  \Delta_1 = \Delta_2 = ... = \Delta_{N-1} = - \tD_N.
\label{LHVb2}
\ee
The equalities can be extended to include $\Delta_N$ by adding two more operators,
\be
  O_N = X\hbox{...}X\tY Y  \hskip1truecm  \hbox{and}  \hskip1truecm O_{N+1} = YX\hbox{...}X\tY X,
\label{ConOps2b}
\ee
from whose ratios we deduce that all of the $\Delta$s are equal,
\be 
  \Delta_k = \cyk - \cxk \equiv \Delta,
\label{LHVb3}
\ee
that is, the HV variations are uniform over qudits.   A similar deduction for the values of individual 
$\cxk$ and $\cyk$ is of course impossible, since these must be random.

To find GHZ contradictions, we may choose $\phio = 2\pi/Nd$, so that $Y^{\otimes N}$ appears at 
$\nu = 1$ and the quantum prediction for its measured value is $Y^{\otimes N} \rightarrow \om$.  The 
hidden variables prediction based on \Eq{LHVb3} is
\be
  v(Y^{\otimes N}) = v(X^{\otimes N})\om^{\sum_k \Delta_k} = \om^{N \Delta}.
\label{LHVb4}
\ee
Consistency with the quantum prediction requires that
\be
  N \Delta = 1,
\label{LHVb5}
\ee
which cannot be satisfied for $N$ equal to any multiple of $d$, or of its factors (excluding unity).   Note 
that $N=2$, which admits a hidden variable construction of perfect correlations \cite{Ngeq3}, is ruled 
out by construction (\Eqs{ConOps2a}{ConOps2b}).   We have thus derived GHZ contradictions for those 
values of  $N$ which elude method 1, down to the smallest factors of $d$, as shown in green in Fig. 2.   

The derivation described above requires $N+3$ observables, composed of three measurement bases 
for two of the qudits, and two measurement bases for remaining qudits.  In comparison, method 1 
required only $N+1$ observables, with two measurement bases for every qudit.   These requirements 
may be taken as the operational definitions of regimes 1 and 2.  Cases for which both methods work 
are assigned to regime 1.


\vskip0.3truecm
\centerline{{\bf Method 3: All $N \geq 3$}}
\vskip0.2truecm

Here we address the remaining cases colored blue in Fig. 2.  These require still further concurrent 
operators, and correspondingly further one-qudit measurement bases, whose numbers will be 
estimated later.

Let us illustrate this method for the most challenging case, $N=3$, from which it will be clear 
how to generalize.   Concurrent operators at $\nu=0$ from the previous section are
\be
  YX\tY,~XY\tY,~X\tY Y,~Y\tY X.
\label{ConOps3a}
\ee
Measured values of unity allow the inference that $\Delta_1 = \Delta_2 = \Delta_3 \equiv \Delta$.   
The necessary additional concurrent operators are built with one-qudit factors defined at multiples of 
the basic angle $\phio$.  We write these as $\YY{n} \equiv X(n\phio)$, for $n = \pm 1, \pm2, ...$, and 
then define the Fibonacci-like sequence of operators, all at $\nu=0$,
\be
    \YY{\pm 2}\YY{\mp 1}\YY{\mp1},~\YY{\mp 2}\YY{\pm 3}\YY{\mp1},~\YY{\mp 2}\YY{\mp3}\YY{\pm 5},
    ~\YY{\pm 8}\YY{\mp 3}\YY{\mp 5},~...,
 \label{ConOps3c}
\ee
from which the necessary operators can be chosen.   Writing the usual hidden variable parameters 
for the new one-qudit factors as 
\be
  \hbox{ln}_{\om}[v(\YY{n}_k)/v(X_k)] = \cyk^{(n)} - \cxk \equiv \Del{n}_k,
\label{LHV3a}
\ee
we can make the following inferences from measured values of unity on operators in 
(\ref{ConOps3c}):   From the first pair, $\Del{2}_1 = 2 \Delta = - \Del{-2}_1$; from the second pair, 
$\Del{3}_2 = 3 \Delta = - \Del{-3}_2$; and so on.   We can stop as soon as the indices add up to 
$d$:   If $d=5$, then just two elements from the sequence, $\YY{-2}YY$ and $\YY{-2}\YY{3}\YY{-1}$, 
suffice to build the hidden variable prediction for $Y\YY{3}Y$, which is  
\be
   v(Y\YY{3}Y) = v(\BX) \om^{5\Delta} = 1.
\label{LHV3b}
\ee
For the quantum prediction, we can choose $\phio = 2\pi/25$, which places  $Y\YY{3}Y$ at  $\nu = 1$, 
with eigenvalue $\om$, and forms a GHZ contradiction.  We have obtained this contradiction with a 
total of 8 concurrent operators.  In the case of $d=7$, one requires four elements from the sequence 
(those containing $\YY{\pm 2}_1$, $\YY{-3}_2$, and $\YY{5}_3$) to build $YY\YY{5}$ for comparison 
with $\BX$.  The hidden variable and quantum predictions,  with $\phio = 2\pi/49$ for the latter, form a 
similar GHZ contradiction with a total of 10 concurrent operators.   For larger $d$, the choice
$\phio = 2\pi/d^2$ remains appropriate.

Contradictions with larger values of $N$ are straightforward extensions:    We employ $N+2$ 
concurrent operators of method 2 (\Eqs{ConOps2a}{ConOps2b} and $\BX$), and select additional 
operators from a similar Fibonacci-like sequence whose number will increase with $d$ but decrease 
with $N$.  The largest  $N$ values for which this method is useful are $N = d-1$.  In these cases, only 
a single operator is required from the sequence, for example $Y^{(2)}X...X\tY\tY$, allowing a hidden 
variable prediction of the $\nu=1$ operator, $Y^{(2)}Y...Y$, to demonstrate the contradiction with a 
total of only $N+4$ concurrent operators.  The choice of $\phio = 2\pi/d^2$ remains appropriate for all 
relevant $N$ ($3 \leq N < d$). 

Note that the above inferences $\Delta_k^{(n)} = n \Delta$ involving multiples of $\phio$ are particle 
{\it specific}, a consequence of minimizing the number of required measurements.   In principle these
inferences could be extended to all qudits by including further operators in the sequence, implying
uniformity of variations as well as linearity as a function of the discrete angle, $n \phio$.
This goes beyond what we need for the desired GHZ contradictions, but, of course, falls short of
of the proof $\Delta \cx_k(\phi) = 0$ based on a continuum of angles.

\vskip.3truecm
\centerline{{\bf Compared Requirements}}
\vskip.2truecm

Let us compare the three regimes with respect to the minimal requirements for each method of 
proof described above.  For each method (1, 2, 3), the minimum number of concurrent $N$-body 
operators  is $N+1$, $N+3$, and $\geq N+4$, respectively.  The corresponding number of 
one-qudit measurement bases is $2^N, 2^{N-2}3^2$, and $(\geq 2)^{N-3}(\geq 3)^3$, where 
superscripts denote the number of particles to which the basis number applies.  Specifically, 
method two requires three bases for two qudits and two bases for all others, while method three 
requires at least three bases for three qudits and at least two for all others.   In method 3, as the
examples show, the number of operators and measurement bases depend on both $d$ and $N$,
generally as increasing functions of $d$ and decreasing functions of $N$.

Note that as we progress through the three methods, the possible inferences are expanded, whether
or not one chooses to make use of them in the proofs.   One can easily see that the concurrent operators 
employed in method 1 allow the inference that some or all of the variations $\Delta_k$ are equal to one 
another \cite{inference}.   With method 2, we infer the uniformity of both $\Delta_k$ and $\tD_k$.   With 
method 3 we infer in addition the linearity ($\Delta_k^{(n)} = n \Delta_k$) for some,  but not necessarily 
all of the qudits $k$. 

\vskip.3truecm
\centerline{{\bf Irreducibility of Contradictions}}
\vskip.2truecm

An $N$-particle contradiction is irreducible if no single qudit can be removed without spoiling it.  Those 
of method 1 were shown to be irreducible in \Ref{RLZL}.   Those of method 2 are clearly irreducible 
because every qudit has both $X$ and $Y$ factors from at least one of the concurrent operators,  all of
which are essential to the proof.   As for method 3, it is difficult to argue in general because of its inherent
flexibility.    But it is easy to see that all of the given examples are irreducible.  It seems plausible that if 
one minimizes the number of operators used in the proof, it will also be irreducible with respect to $N$.
  
A contradiction is genuinely $d$-dimensional if the $d \times d$ matrices $X_k$ and $Y_k$ cannot 
be simultaneously block-diagonalized, or equivalently, if one cannot find an eigenstate of $X_k$, 
and another of $Y_k$, with vanishing inner product \cite{RLZL}.   Fig. 1 (a and b) provides an 
elegant general proof that indeed one cannot:  Note that \Eq{RotPsi} applies to the $N=1$ case, 
where the rotated states (a) are eigenstates of the rotated $X_k(\phik)$ matrix (b).  Covariance
shows that the states at the special points $2\pi\nu/d$ form a basis of eigenstates of $X_k$, while 
those at points $\phio+2\pi\nu/d$ form a basis for $Y_k$.   Equation \ref{inner} shows that the inner 
product of any pair of states, one from each basis, does not vanish.  Since both bases are 
nondegenerate, there is no other choice of eigenstates.    So our contradictions are genuinely 
$d$-dimensional.   This argument applies equally well to other one-qudit factors $\YY{n}$.

\vskip0.3truecm
\centerline{{\bf Physical Interpretation}}
\vskip0.2truecm

As we have seen, the requirement for GHZ contradictions is the multiplicity of $N$-body operators 
that share a common eigenstate, with two or more differing eigenvalues.  
To relate this to rotational covariance, note that if we rotate $\ket{\Psi}$ through an angle ($-2\pi/d$), 
it is still an eigenstate of $\BX$, but with eigenvalue $\om$ rather than 1.  Equivalently, if we
rotate $\BX$ through ($+2\pi/d$), it still has $\ket{\Psi}$ as an eigenstate, but with eigenvalue 
$\om$.  The difference is that in the latter case, there many operators arising 
from the many ways of distributing the net rotation among the factors.    All of these 
rotated operators correspond to the same equivalently rotated state, which is oblivious to the
distribution.  Thus, the multiplicity arises from the invariance of the GHZ state under the 
$\Phi$-preserving rotations that relate all of the operators at ($+2\pi/d$).

The GHZ contradictions require in addition that operators at different points (separated by $2\pi/d$
or a multiple) have different eigenvalues in the same state, a property of GHZ covariance.  The
common feature of all successful proofs is that HV functions $[\cxk(\phi)]$ are so constrained by 
the quantum predictions at any one point, that they cannot reproduce those at another point.

For a broader perspective, let us return to the discussion of continuous transformation properties 
of Sec. II, where we showed that states transform according to the circle group, $\CT$, while 
operators transform as $\CTN$.   We show here that the expectation value, 
$\ave{\BRphik \BX \BRinvphik}$, transforms simply as $\CT$ under {\it relative} rotations ($\Phi$) 
between the operator and the state. Note first that the one-body operator, \Eq{rotXk1}, and hence 
its expectation value in the fixed one-particle state, \Eq{GHZ1}, transform as two-dimensional 
representations of $\CT$.  Second, the arguments leading to \Eq{rotBX3} show that the expectation 
values of $N$-body operators $[\BXPhi]$ in fixed $N$-particle GHZ states (\ref{GHZ1}) are given by 
the same function of $\Phi$, independent of $N$. This result generalizes to the probability distribution 
of measurement outcomes for rotated $N$-body operators in fixed GHZ states, whether these 
outcomes are definite or probabilistic.  In this respect, the $N$-body outcomes are reduced to 
one-body outcomes. 

\section{Conclusions}

In summary, we have shown that the many-particle rotational symmetry characterizing GHZ states 
cannot be satisfied by hidden variables for $N \geq 3$ particles of any dimension $d$.  A discrete 
subset of symmetry operations involving net rotations of $2\pi/d$ identifies concurrent operator sets 
that exhibit specific, experimentally verifyable GHZ contradictions that, in total, are similarly
unrestricted.    These contradictions fall into three regimes defined according to increasing 
numbers of $N$-body operators as well as one-qudit measurement bases required for theoretical 
proofs and experimental tests.  The first regime recovers existing proofs \cite{LLK,RLZL}, the second 
regime adds new proofs to some existing ones \cite{subsets}, and the third regime consists entirely
of new proofs that complete the catalog of all possible $N$ values for every $d$.  

The current results are interesting in part because of the novelty of the odd-$d$ contradictions  (of 
\Ref{RLZL} as well as the present work), which demonstrate that the concurrent operator approach 
places almost all dimensions on equal footing with respect to the existence of GHZ contradictions.  
State-dependent contradictions exist for all $d$, and state-independent for none, with the exception 
of $d=2$, where Pauli operators are recovered by appropriate rotations.  In contrast, there appears 
to be a fundamental distinction between even and odd dimensions when applying stabilizer sets 
in higher dimensions, in that state-independent contradictions have been found for all even 
$d$ \cite{CMP,TYO13,WA12}, but shown not to exist for any odd $d$ \cite{other}.  A limited number 
of state-dependent contradictions have been found for some odd $d$ \cite{Howard}.

The successful use of GHZ symmetry in the present work raises the question of a more general 
relationship between entangled-state symmetries and GHZ contradictions.   Will such particular
symmetries more generally favor concurrent operator sets over stabilizer sets?   If so, then since 
entanglement and nonlocality are useful resources for quantum information processing, perhaps 
concurrent operators will prove useful as well.

\begin{thebibliography}{99}
%

\bibitem{GHZ} D.M. Greenberger, M.A. Horne, and A. Zeilinger, in 
   {\it Bell's Theorem, Quantum Theory and Conceptions of the Universe},
   edited by M. Kafatos (Kluwer Academic, Dordrecht, 1989), p. 69, and
   eprint arXiv:quant-ph/0712.0921.
\bibitem{GHZS} D.M. Greenberger, M.A. Horne, A. Shimony,  and A. Zeilinger,
   \source{Am. J. Phys.}{58}{1131}{1990}.
\bibitem{HR83}  P. Heywood and M.L.G. Redhead, \source{Found. Phys.}{13}{481}{1983} 
    found the first nonprobabilistic contradictions, based on the singlet state of two spin-1 
    particles.
\bibitem{Mermin1} N.D. Mermin, {\it Phys. Rev. Lett.}{\bf 65}, 3373 (1990).
   Also see N.D. Mermin, {\it Rev. Mod. Phys.} {\bf 65}, 803 (1993).
\bibitem{Bell1966}J. S. Bell, \source{Rev. Mod. Phys.}{38}{447}{1966}. 
\bibitem{KS} S. Kochen and E. P. Specker, \source{J. Math. Mech.}{17}{59}
   {1967}.
\bibitem{Bell1964} J. S. Bell, \source{Physics}{1}{195}{1964}.
\bibitem{Mermin2} N. D. Mermin, {\it Phys. Rev. Lett.} {\bf 65}, 1838 (1990).
\bibitem{Ardehali} M. Ardehali, \source{Phys, Rev. A}{46}{5375}{1992}. 
\bibitem{Hardy} L. Hardy, \source{Phys. Rev. Lett.}{71}{1665}{1993}.
\bibitem{KP} M. Kernaghan and A. Peres, \source{Phys. Lett. A}{198}{1}{1995}.
\bibitem{Cabello96} A. Cabello, J.M. Esteberanz, G. Garcia-Alcaine, 
   \source{Phys. Lett. A}{212}{183}{1996}.
\bibitem{Lisonek} P. Lisonek, P. Badziag, J.R. Portillo, and A. Cabello, 
    eprint arXiv:1308.6012v1 [quant/ph].
\bibitem{Kas.et.al.} D. Kaszlikowski et. al. \source{Phys. Rev. Letters}{85}{4418}{2000}.
\bibitem{Collins} D. Collins et. al. \source{Phys. Rev. Letters}{88}{040404}{2002}.
\bibitem{Fu} Li-Bin Fu, \source{Phys. Rev. Lett.}{92}{130404}{2004}.
\bibitem{Klyachko} A.A. Klyachko, M.A. Can, S. Binicioglu, and A.S. Shumovsky, 
    \source{Phys. Rev. Lett.}{101}{020403}{2008}.
\bibitem{Cabello08} A. Cabello, \source{Phys. Rev. Letters}{101}{210401}{2008}.
\bibitem{Badziag} P. Badziag, I. Bengtsson, A. Cabello, and I. Pitowsky, 
    \source{Phys. Rev. Letters}{103}{050401}{2009}.
\bibitem{QSS} M. Hillery, V. Bu\v{z}ek, and A. Berthiaume, \source{Phys. Rev. A}{59}
    {1829}{1999}; and R. Cleve, D. Gottesman, and H.-K. Lo, \source{Phys. Rev. Letters}{83}
    {648}{1999}. 
\bibitem{SG01} V. Scarani and N. Gisin, \source{Phys. Rev. A}{65}{012311}{2001}.
\bibitem{DiVP} D. DiVincenzo and A. Peres , \source{Phys. Rev. A}{55}{4089}{1997}. 
\bibitem{Frowis} F. Fr\"{o}wis and W. D\"{u}r, \source{Phys. Rev. Lett.}{106}{110402}
   {2011}. 
\bibitem{ZK99} M. Zukowski and D. Kaszlikowski, \source{Phys. Rev. A}
   {59}{3200}{1999}.
\bibitem{KZ02} D. Kaszlikowski and M. Zukowski, \source{Phys. Rev. A}
   {66}{042107}{2002}.
\bibitem{CMP} N.J. Cerf, S. Massar and S. Pironio, {\it Phys. Rev. Lett.}
   {\bf 89}, 080402 (2002).  
\bibitem{LLK} J. Lee, S.-W. Lee, and M.S. Kim, \source{Phys. Rev. A}{73}
   {032316}{2006}.
\bibitem{EPR} A. Einstein, B. Podolsky, and N. Rosen, \source{Phys. Rev.}
   {47}{777}{1935}.
\bibitem{TYO13}  W. Tang, S. Yu, and C.H. Oh, \source{Phys. Rev. Lett.}
   {110}{100403}{2013}.
\bibitem{TYO13b} W. Tang, S. Yu, and C.H. Oh, eprint arXiv:1303.6740v1 [quant/ph].
\bibitem{WA13} M. Waegell and P.K. Aravind, \source{Phys. Rev. A}{88}{012102}{2013};
  see also \source{J. Phys. A: Math. Theor.}{44}{505303}{2011}; and {\bf 45}, 405301
  (2012).
\bibitem{WA12} M. Waegell and P.K. Aravind, \source{Phys. Lett. A}{377}{546}{2013}.
\bibitem{RLZL} J. Ryu, C. Lee, M. Zukowski, and J. Lee, \source{Phys. Rev. A}{88}
   {042101}{2013}.
\bibitem{RLZL2}  J. Ryu, C. Lee, Y. Zhi, R. Rahaman, D.G. Angelakis, J. Lee, and 
    M. Zukowski, eprint arXiv:1303.7222v2 [quant/ph].
\bibitem{Howard} M. Howard, E. Brennan, and J. Vala, \source{Entropy}{15}{2340}{2013}.  
\bibitem{other} \Ref{Howard} shows how this conclusion emerges from results
    of D. Gross, \source{J. Math. Phys.}{47}{122107}{2006}; and V. Veitch, C. Ferrie, 
    D. Gross, and J. Emerson, \source{New J. Phys.}{14}{113011}{2012}.
\bibitem{Ngeq3} The extrapolations of GHZ states to $N=2$ are Bell states, and it is well
    known \cite{Bell1964} that hidden variables succeed in describing the perfect 
    correlations observed for two qubits.   Nonprobabilistic contradictions do exist, 
    nontheless, for the singlet state of two qutrits \cite{HR83}.
\bibitem{subsets} Referring to the green squares in Fig. 2, \Ref{TYO13} proved the 
   even-$d$ cases, and \Ref{RLZL2} proved odd-$N$ cases below the line $N=d+1$.  
   These and remaining cases are covered by the alternative method described in the
   text.
\bibitem{inference} If $f$ and $N$ are coprime, then one may infer that all $\Delta_k$ 
   are equal.  If not, then uniformity can still be inferred within subsets.
%
%
\end{thebibliography}
\end{document}